\documentclass[showpacs,reprint,amsmath,amssymb,superscriptaddress,aps,prl]{revtex4-1}
\usepackage[percent]{overpic}
\usepackage{amsmath,amssymb}
\usepackage{graphicx}
\usepackage{float}
\usepackage{xcolor}

\begin{document}

\title{All optical implementation of a time-domain ptychographic pulse reconstruction set-up}
\author{Dirk-Mathys Spangenberg}\email{Corresponding author: dspan@sun.ac.za}
\affiliation{LRI, Stellenbosch University, Private Bag X1, 7602 Matieland, South Africa}
\author{Michael Br\"ugmann}
\affiliation{IAP, University of Bern, Sidlerstr. 5, 3012 Bern, Switzerland}
\author{Erich Rohwer}
\affiliation{LRI, Stellenbosch University, Private Bag X1, 7602 Matieland, South Africa}
\author{Thomas Feurer}
\affiliation{IAP, University of Bern, Sidlerstr. 5, 3012 Bern, Switzerland}


\begin{abstract}
An all optical implementation of pulse reconstruction using time-domain ptychography is demonstrated showing excellent results. Setup and reconstruction are easy to implement and a number of drawbacks found in other second order techniques are removed, such as the beam splitter modifying the pulse under consideration, the time ambiguity, or the strict correspondence between time delay increment and temporal resolution. Ptychography generally performs superior to algorithms based on general projections, requires considerable less computational effort and is much less susceptible to noise.
\end{abstract}



\maketitle

\section{Introduction}

Since the discovery of ultrashort laser pulses, characterization of these has been an imminent task as no detector exists which can characterize the temporal intensity of such pulses directly. In 1967, Weber demonstrated that a nonlinear auto-correlation scheme can be employed to estimate the pulse duration \cite{Weber_1967}. Such an arrangement is now classified as a second-order background-free intensity auto-correlation measurement. About 20 years later, Diels and co-workers introduced the second-order interferometric auto-correlation with the aim to not only estimate the pulse duration but to fully characterize the electric field of ultrashort laser pulses, e.g., their spectral amplitude and phase \cite{Diels_1985}. In the following years, several further ingenious schemes were proposed and demonstrated, such as FROG \cite{trebino-2000}, STRUT \cite{Chilla-1991}, SPIDER \cite{Iaconis1998}, PICASO \cite{Nicholson_1999}, MIIPS \cite{Xu_2006}, and many variations thereof.

Recently, we migrated a spatial lens-less imaging technique, ptychography, from the spatial domain to the temporal domain. Ptychography is related to the solution of the phase problem in crystallography and was first proposed by Hoppe \cite{Hoppe-1969}. In ptychography, the real space image of an object, in particular its amplitude and phase, is reconstructed iteratively from a series of far-field diffraction measurements. Each of those is recorded after either moving the object or the coherent illumination beam in a plane perpendicular to the propagation direction of the illumination beam. The transverse shift of the illumination beam is smaller than its spatial support, so that subsequent far-field diffraction patterns result from different, but overlapping, regions of the real space object. The spatial resolution of the reconstructed image is determined by the angular range of scattered wave vectors that can be recorded with a sufficiently high signal-to-noise ratio, by the positioning accuracy, and by the stability of the entire set-up. While the original procedure relies on the illumination beam being fully characterized \cite{McCallum_1992}, the substantial redundancy in the measurement also allows for reconstruction of the illumination beam \cite{Thibault_2009,Maiden_2009}.

We demonstrated that both ptychography modalities can be adapted to time domain imaging \cite{Spangenberg_2015a,Lucchini2015}. We reconstructed 'objects' uniform in space but varying with time by illuminating the objects with a sequence of known partially overlapping, time delayed coherent probe pulses. For each time delay a far-field diffraction pattern, i.e. wavelength resolved spectrum, was recorded . Time domain ptychography then reconstructs the amplitude and the phase of the object from the resulting spectrograms. We also showed that the extended ptychography algorithm can characterize the object as well as the  illumination pulse. The object was an attosecond XUV pulse and the illumination pulse an IR streak pulse, both unknown. Finally we demonstrated that a slightly modified reconstruction algorithm can be used for pulse characterization purposes \cite{Spangenberg_2015b}. The experimental demonstration, however, was done with a setup containing two spatial light modulators (SLM) in order to have the most flexibility on the pulse modulation. Here we present a simple experimental setup, without a SLM, which is exclusively dedicated to ultrafast pulse characterization, which we label PIRANA. Compared to other methods the setup and reconstruction are easy to implement and a number of shortcomings are removed, such as the beam splitter modifying the pulse under consideration, the time ambiguity, or the strict correspondence between time delay increment and temporal resolution.

\section{Theory}

The ptychographic scheme differs from other pulse reconstruction modalities, such as the principle component generalized projections algorithm (PCGPA) \cite{Kane_1999}, specifically related to the following: 1) The time delay increment is not related to the desired temporal resolution or the wavelength sampling of the spectrometer, but only to the duration of the illumination pulse. 2) Typically, only a few spectra have to be recorded. 3) The small number of spectra to process and the robust algorithm result in an extremely fast convergence of the retrieval algorithm.

The PCGPA algorithm relies on a spectrogram sampled on an $M \times M$ grid that satisfies the sampling condition $\delta\nu \delta t = 1/M$ where $\delta \nu$ is the angular frequency step size and $\delta t$ is the time steps size. Typically the spectral axis has a higher sampling rate than the temporal axis, and as a consequence the temporal axis needs to be interpolated. Naturally, interpolation will neglect any temporal structure finer than the original sampling, thus PCGPA cannot accurately reconstruct temporal features which vary more rapidly than the time delay increment. In contrast, the ptychographic scheme has no link between temporal sampling and time delay. Time domain ptychography operates on two sampling grids which are largely independent from another. The object and the illumination pulse are sampled on an equidistant temporal grid, with $M$ samples equally spaced by $\delta t$, which is determined by the resolution and total spectral range of the spectrometer used. The second grid is that of the time delays and consists of $N$ samples equally spaced by $\delta\tau$. If both grids span the same time window, their frequency increments $\delta\nu$ are identical and we find $\delta\nu \delta t = 1/M$ and $\delta\nu \delta\tau = 1/N$, respectively. The only constraint on the two integers is $N \leq M$ but typically $N$ is orders of magnitudes smaller than $M$.

The sampling requirements for ptychography is simpler. Ptychography requires $N$ spectra $I_n(\nu)$ which are recorded at different time delays $t_n$ ($n = 1 \ldots N$) between the object and the illumination pulse. All spectra combined result in a spectrogram $S(\omega,\tau)$ sampled on an $M \times N$ grid. In ptychography the time delay increment $\delta\tau$ is related only to the duration of the slowly varying envelope of the illumination pulse. The relevant quantity is the fundamental sampling ratio $R$. It is defined as the ratio of the full widths at half maximum duration (FWHM) of the illumination pulse over the time delay increment and if both are identical the fundamental sampling ratio is equal to one. For a fundamental sampling ratio $R > 1$ the illumination pulse overlaps with parts of the object several times and this overlap increases the redundancy in the data recorded. It is well known that this redundancy can be used to not only reconstruct the object but also the illumination pulse \cite{Thibault_2009,Maiden_2009}.

For the purpose of ultrafast pulse reconstruction we identify the 'object' with the pulse $E(t)$ to be characterized and the illumination pulse, henceforth labeled gate $P(t)$, with a filtered copy of the same pulse, i.e. $P(t) = E(t) \otimes H(t)$. The spectral filter is characterized by its linear transfer function $H(\omega)$, $H(t)$ is its Fourier transform and $\otimes$ denotes a convolution. Since the gate is derived from the pulse to be characterized we expect ptychographic pulse reconstruction to work for $R > 1$, that is we exploit the redundancy to reconstruct the pulse as well as the gate. The spectral filter we use is a bandpass centered at the carrier frequency of the laser with a variable width. The idea is that narrowing the spectrum results in an illumination pulse which is longer than the object.

As a starting point for the reconstruction algorithm we assume the initial estimate of the pulse to be identical to the spectral filter, i.e. $E_{j=1,n=0}(t) = H(t)$. This forces the position of solution and gate to be located around time zero. This prevents the solution from jumping around until most of the energy is arbitrarily located at some temporal position after which the algorithm reconstructs both gate and pulse as if they are located there. The reason for this temporal position ambiguity is due to the gate being derived from the pulse itself. In every iteration $j$ all measured spectra ($n = 1 \ldots N$) are processed. For ascending $n$ the algorithm first updates the current estimate of the gate and hereafter the estimate of the pulse. The gate then is derived from the pulse according to

\begin{equation}
\label{eq1}
P_{j,n}(t-\tau_n) = \mathcal{F}^{-1}\left[ E_{j,n}(\omega) H(\omega) \textrm{e}^{i (\omega-\omega_0) \tau_n} \right].
\end{equation}

Updating $P_{j,n}(t)$ is not part of the original ptychographic reconstruction algorithm, however is crucial to achieve convergence. Next, we calculate the exit field $\xi_{j,n}(t,\tau_n)$ for a particular time delay $\tau_n$ between the previous estimate of the pulse $E_{j,n-1}(t)$ and the gate $P_{j,n}(t-\tau_n)$. 

\begin{equation}
\label{eq2}
\xi_{j,n}(t) = E_{j,n-1}(t) P_{j,n}(t-\tau_n)
\end{equation}

From $\xi_{j,n}(t)$ we calculate the Fourier transform $\xi_{j,n}(\omega)$ and replace its modulus by the square root of the corresponding spectrum $S_n(\omega)$ while preserving its phase. After an inverse Fourier transformation, the new function $\xi'_{j,n}(t)$ differs from the initial estimate, and the difference is used to update the current estimate of the pulse

\begin{equation}
\label{eq3}
E_{j,n}(t) = E_{j,n-1} (t) + \beta \; U_{j,n}(t-\tau_n) \left[ \xi'_{j,n}(t) - \xi_{j,n}(t) \right]
\end{equation}

with the weight or window function

\begin{equation}
\label{eq4}
U_{j,n}(t) = \frac{P_{j,n}(t)}{\textrm{max}(|P_{j,n}(t)|)} \; \frac{P^*_{j,n}(t)}{|P_{j,n}(t)|^2+\alpha}
\end{equation}

and the two constants $\alpha < 1$ and $\beta \in [0…1]$. Before proceeding with the next iteration $j+1$, we set $E_{j+1,n=0}(t) = E_{j,n=N}(t)$. We quantify the reconstruction outcome through the root mean square (rms) of the deviation between measured and reconstructed spectrogram. The rms value serves as a quality factor for the success of reconstruction. 

\begin{equation}
\label{eq5}
\mathrm{rms} = \sqrt{\frac{\sum \left[ S_\mathrm{exp}(\omega,\tau) - S_\mathrm{rec}(\omega,\tau) \right]^2}{N M}}
\end{equation}

\section{Simulations}

\begin{figure}[htb]
\centering
\includegraphics[width=85mm]{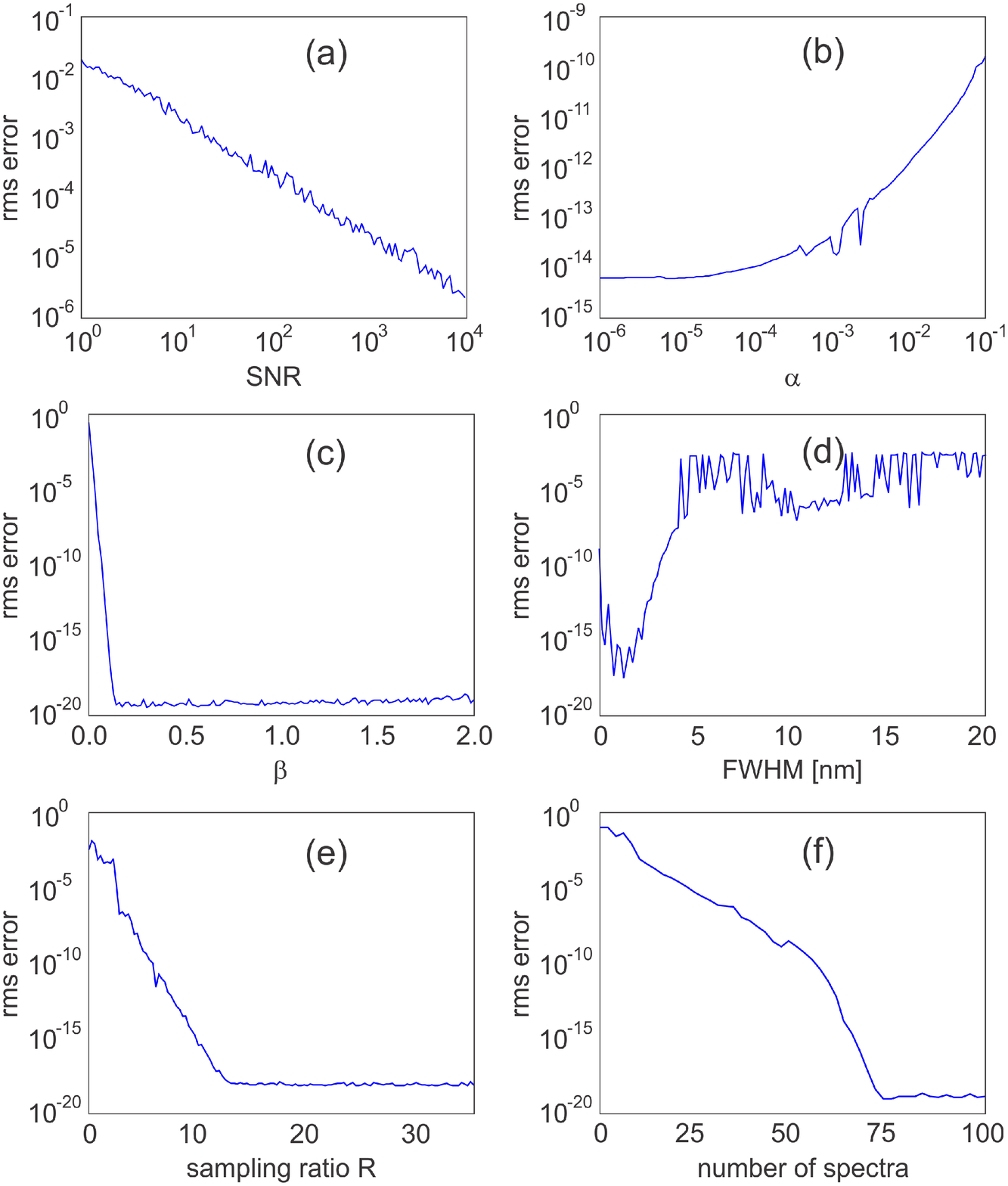}
\caption{Reconstruction error for a chirped pulse; pulse duration 80~fs, linear chirp $-10000$~fs$^2$. (a) rms versus signal-to-noise ratio, (b) $\alpha$, (c) $\beta$, (d) FWHM of spectral bandpass, (e) sampling ratio $R$, and number of spectra $N$. The parameters, except for the one varied, are fixed to $\alpha = 10^{-3}$, $\beta = 0.25$, FWHM = 3~nm, sampling ratio $R = 8$ and number of spectra $N = 49$. For each parameter set we performed 500 iterations.}
\label{fig_rms}
\end{figure}

It may appear somewhat challenging to identify suitable values for the width of the spectral bandpass filter, the time delay steps, or the weighting constants $\alpha$ and $\beta$ since no theoretical framework exists which can hint to their optimal values. In this section, we therefore reconstruct a chirped pulse (pulse duration 80~fs, linear chirp $-10000$~fs$^2$) while varying all relevant parameters, namely $\alpha$, $\beta$, $R$, the FWHM of the spectral bandpass filter, and the number of recorded spectra $N$. The results are summarized in Fig.~\ref{fig_rms}. The reconstruction parameters, except for the one varied, were fixed to $\alpha = 10^{-3}$, $\beta = 0.25$, bandpass FWHM = 3~nm, sampling ratio $R = 8$ and number of spectra $N = 49$. For each set of parameters we performed 500 iterations.

Figure~\ref{fig_rms}a) shows the rms error as a function of the signal-to-noise ratio (SNR). For this we added white noise with an amplitude given by the signal-to-noise ratio to the simulated spectrogram prior to reconstruction. The linear decrease in the double logarithmic plot indicates that the rms error is solely dominated by the noise present in the simulated spectrogram and not by the quality of the reconstruction. Only for SNR values below 5 do we see an influence of the decreasing quality of the reconstructed spectrogram. A similar behavior was already observed in the ptychographic reconstruction of attosecond pulses \cite{Lucchini2015} where a SNR value as low as 3 was sufficient for successful reconstruction. Henceforth we abstain from adding noise to the spectrogram and vary only the reconstruction parameters in order not to obscure their influence. In several cases the reconstruction error seems to level off around $10^{-18}$ which we found is the numerical noise floor. Figure~\ref{fig_rms}b) shows the rms error as a function of $\alpha$ and we find that $\alpha$ has no influence on the reconstruction results. This is to be expected because $\alpha$ is relevant only if the spectrogram is noisy. Next, we vary $\beta$ in the range between 0 and 2 and find perfect reconstruction for $\beta$ larger than 0.15 and up to 2. As expected the bandwidth of the spectral bandpass filter is somewhat more critical. Nevertheless, for all pulses investigated we find that a width of less than one half of the laser's bandwidth always produces excellent reconstruction results. One of the most relevant parameters in standard ptychography is the fundamental sampling ratio $R$ which essentially determines the amount of redundancy in the data recorded. The higher $R$ the more redundancy the data contains and the smaller the rms error becomes until it reaches the numerical noise floor around $R \approx 5$. Lastly, we vary the number of recorded spectra keeping the sampling ration $R$ fixed at a value of 8. We find that as little as 35 spectra result in a excellent quality pulse reconstruction which is in stark contrast to almost all other iterative reconstruction algorithms. The simulations were also performed for a variety of other pulses, i.e. more or less spectral bandwidth, higher order chirps, double pulses, and pulse trains, always with a very similar outcome. In essence all parameters can be varied within a relatively large range without altering the outcome of the reconstruction; mostly only the convergence rate changes. 

\section{Experiment and discussion}

From the theory section above we see that the set-up must in essence implement the spectral intensity measurement

\begin{equation}
\left| \mathcal{F} \left[ E(t-\tau) \; \mathcal{F}^{-1} \left[ E(\omega) H(\omega) \right] \right] \right|^2
\end{equation}

Therefore, conceptually the optical requirement is to split the pulse in two replicas, spectrally filter one of them by $H(\omega)$, multiply the result with the unaltered time delayed second replica and record the generated sum-frequency spectrum. In order to split the pulse into two replicas a grating is used where the zero order gives the unaltered pulse time delayed by a position adjustable mirror and the first order dispersed light is used to generate the spectrally filtered pulse by sending it through a 4f-geometry with a spectral slice filter. After passing the grating a second time, the time delayed and the filtered replica are propagating collinear with a vertical offset. The two pulses are then focused in a beta-bariumborate (BBO) crystal, sufficiently thin as to guarantee perfect phase matching for the entire bandwidth. The resulting sum-frequency light is analyzed by a spectrometer. Using a grating as a beam splitter has a number of advantages. First, the pulse to be characterized does not experience any additional phase modulation as it is the case for a standard beam splitter, and second, the spectral filter can be adjusted and modified easily by selecting a suitable spatial mask.

\begin{figure}[htb]
\centering
\includegraphics[width=65mm]{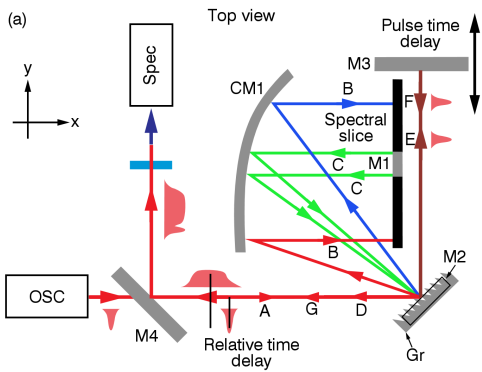}
\caption{Schematic of the ptychographic pulse characterization set-up. The incoming pulse (A) is split by the grating in two diffraction orders (B) and (E). The angular dispersed first order light (B) is focused to a plane mirror from which only a slice of the spectrum (C) is reflected. Its spectral width and center wavelength can be easily varied. The zero order diffraction light (E) is also reflected back to the grating (F), however this mirror is mounted on a motorized translation stage in order to introduce a variable delay time between (C) and (F). After the grating, both orders, i.e. (D) and (G), propagate collinear to the pick-off mirror M4 and are focused in a BBO crystal.}
\label{fig_setup}
\end{figure}

The experimental realization, illustrated in Fig.~\ref{fig_setup} (optical elements denoted by cursive text and pulse paths in brackets), shows the pulsed laser source on the left, a Fusion from Femto Lasers, with the beam passing under the mirror \emph{M4} before it is incident on the grating \emph{Gr}. At the grating \emph{Gr} the zero and the first order reflections generate both the long probe pulse, (beam path ABCD) by sending the first order light through a 4f-geometry consisting of a cylindrical mirror \emph{CM1} with a focal length of 100~mm and a mirror with an opaque cover with a slit cut out \emph{M1} to create a spectral slice filter in the spectral plane, and the unaltered pulse, (beam path AEFG) with a movable mirror \emph{M3} allowing for a variable time delay. The two mirrors \emph{M1} and \emph{M3} are slightly tilted upwards at different angles so that the back-reflected pulses arrive at different heights at the grating. In fact the zero order pulse arrives above the grating where it is reflected from a mirror such that the two pulses (G) and (D) propagate parallel but on top of each other. They are then picked off by mirror M4 and directed and focused to the BBO crystal. The sum-frequency spectrum is recorded with an Avantes spectrometer. The spectrometer has 256~pixel covering the range between 350~nm and 450~nm with a resolution of 0.75~nm. Although we recorded all spectrograms with a time delay increment of 6.67~fs, for the reconstruction we used only every 10th spectrum giving a $\delta\tau$ of 66.7~fs. The spectral bandpass had the same center frequency as the laser and was measured to be approximately Gauss-shaped with a FWHM of 3~nm. As a result the gate had a minimum duration of about 300~fs, resulting in a fundamental sampling ratio $R$ between 4 and 5. The other reconstruction parameters were $\alpha=0.1$, $\beta=0.4$ and 350 iterations.

In the experiment we verify the functionality of the set-up with two sets of measurements. The first compares the theoretically calculated spectral phase resulting from propagation through glass with values obtained from reconstruction and the second set compares measurement and reconstruction of two double pulse waveforms with different temporal spacing between the two sub-pulses.

\begin{figure}[ht]
\centering   
\includegraphics[width=\columnwidth]{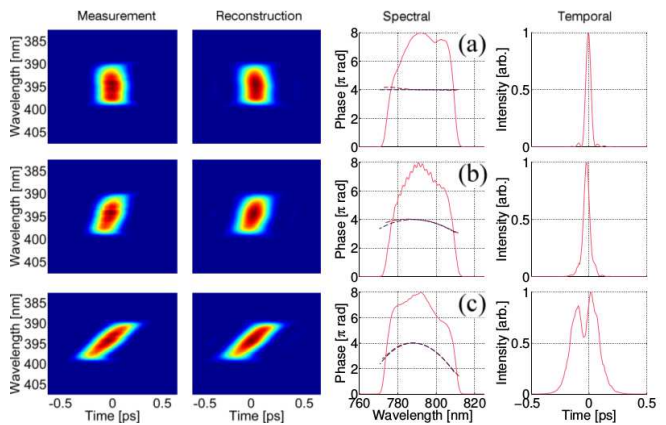}
\caption{Measured spectrogram and reconstruction results of an input pulse with (a) no glass, (b) a medium glass block (30~mm BK7) and (c) a long glass block (100~mm fused silica) inserted in the beam path.}
\label{fig_3_glass}
\end{figure}

In the first set we took three measurements, i.e. a reference measurement with no glass in the beam path, with a medium thickness piece of glass (30~mm of BK7) and finally with a long piece of glass (100~mm of fused silica). The oscillator was tuned to a center wavelength of 788~nm with a spectral bandwidth of 35~nm. In each of the three cases a spectrogram was measured and the spectral amplitudes and phases were reconstructed. In Fig.~\ref{fig_3_glass}, column one and two, we show the measured and reconstructed spectrogram for each case. The agreement between the two is excellent and we find an rms error of $9.3 \cdot 10^{-4}$, $1.5 \cdot 10^{-3}$ and $1.3 \cdot 10^{-3}$, respectively. In the third column we show the reconstructed spectral amplitude and phase in red. The reconstructed phase is compared to the theoretically calculated phase in blue (on-line version) assuming the known dispersion parameters of BK7 and fused silica. We find excellent agreement between the reconstructed and the expected phase values for 30~mm BK7 as well as for 100~mm fused silica. Lastly, column four displays the reconstructed temporal intensity. In order to further quantify the reconstruction results we extract the quadratic spectral phase coefficients from a polynomial fit to the reconstructed and the calculated phase. The quadratic spectral phases were found to agree excellently if we subtract the quadratic spectral phase present when there is no glass inserted in the system (reference measurement). The reconstructed values are compared to their theoretical counterparts in Table~\ref{tab1}.

\begin{table}[htb]
\centering
\caption{\label{tab1} Column one, glass inserted into the system; column two, the theoretical quadratic spectral phase coefficient in fs$^2$ introduced by such a piece of glass for a central wavelength of 788~nm; column three, the value extracted from the reconstructed phase in fs$^2$; column four, the ratio of theoretical over reconstructed quadratic phase coefficient.}
\begin{tabular}{l|c|c|c}
\hline
Material  & Theory  & Retrieved & ratio \\
\hline
No glass                    &              0 &   182.4 $\pm$ 0.4 &     --- \\
30~mm BK7               &     -687.1 &   -654.8 $\pm$ 3.9 & 1.05 \\
100~mm fused silica &   -1857.5 & -1834.2 $\pm$ 0.4 & 1.01 \\
\hline
\end{tabular}
\end{table}

\begin{figure}[htb]
\centering   
\includegraphics[width=\columnwidth]{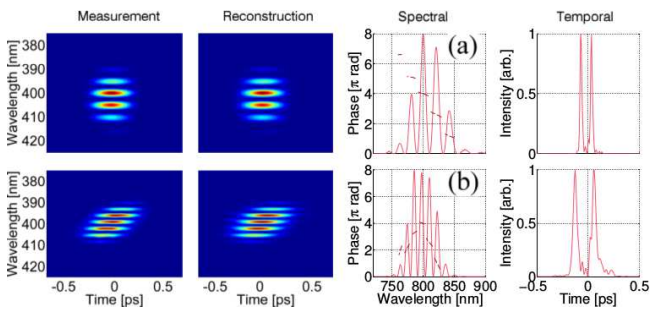}
\caption{Measured spectrogram and reconstruction of two double pulses with (a) no glass and (b) a medium thickness piece of glass (30~mm BK7) inserted in the beam path.} 
\label{fig_dbl_pulse}
\end{figure}

Next we measured and reconstructed double pulses, thus, testing the algorithm for a notoriously difficult problem. The oscillator was tuned to a center wavelength of 788~nm with a spectral bandwidth of 55~nm and was operated in double pulse mode. We show that double pulses can be readily reconstructed for the case when there is no glass inserted in the beam path but also when we introduce additional dispersion to the beam path. The resultant reconstructions are shown in Fig.~\ref{fig_dbl_pulse}.

The top row in Fig.~\ref{fig_dbl_pulse} shows the measured and the reconstructed spectrogram for a double pulse with virtually no chirp. The reconstructed spectral amplitude and phase are shown in the third column and the temporal intensity in the fourth column. The rms error is $1.8 \cdot 10^{-2}$. The reconstructed inter-pulse separation is 103~fs and we observe a $\pi$ phase jump from one spectral peak to the next as expected from theory. The bottom row in Fig.~\ref{fig_dbl_pulse} shows the results for a double pulse with an about two times larger inter-pulse separation which has passed through additional dispersive material. The rms error of the reconstructed spectrogram is $1.0 \cdot 10^{-2}$. The reconstructed inter-pulse separation is 180~fs and the resulting, mainly quadratic phase is easily recognized and superimposed on the $\pi$ phase jumps.

\section{Conclusion}

We demonstrated an all optical implementation of pulse reconstruction using time-domain ptychography with second harmonic generation showing excellent results. Setup and reconstruction are easy to implement and a number of drawbacks found in other techniques are removed. Through simulations we showed that all relevant reconstruction parameters can be varied within a relatively large range influencing only the rate of convergence but not the quality of the final result. We performed a quantitative test by reconstruction of pulses after propagation through known amounts of dispersive material. We also applied the reconstruction to a double pulse which is a notoriously difficult problem for many other algorithms. In all cases we found excellent performance of the ptychographic pulse reconstruction. Moreover the small number of spectra and the redundancy in the data result in a fast and robust reconstruction. In general, ptychography was found to perform superior to algorithms based on general projections, requires considerable less computational effort and is much less susceptible to noise.

\section*{Acknowledgments}
We gratefully acknowledge financial support from the National Research Foundation, the CSIR NLC, and the NCCR MUST research instrument of the Swiss National Science Foundation.


\end{document}